\documentstyle[amsfonts,prb,aps,twocolumn,epsf,floats]{revtex}

\newcommand{\beq}{\begin{equation}}
\newcommand{\eeq}{\end{equation}}
\newcommand{\bea}{\begin{eqnarray}}
\newcommand{\eea}{\end{eqnarray}}

\newcommand{\etal}{{\em et al.}}
\newcommand{\ie}{{i.e.}}

\newcommand{\hs}{\hat{\sigma}}
\newcommand{\hS}{\hat{S}}
\newcommand{\hsi}{\sigma}
\newcommand{\htau}{\hat{\tau}}
\newcommand{\phy}{\left|{\rm phys}\right>}
\newcommand{\hP}{\hat{P}}
\newcommand{\Tr}{{\rm Tr}}
\newcommand{\iD}{i_{\cal D}}
\newcommand{\cD}{{\cal D}}

\def\subsub#1{\noindent{\bf #1:}}
\def\hh#1{\hat{#1}}
\def\jour#1#2#3#4{{#1}{\bf #2}, #3 (#4)}
\def\tit#1#2#3#4#5{{#1}{\bf #2}, #3 (#4)}
\def\jmp{J.\ Math.\ Phys.\ }

\def\rmp{Rev.\ Mod.\ Phys.\ }
\def\npb{Nucl.\ Phys.\ B\ }

\def\plb{Phys.\ Lett.\ B\ }
\def\prep{Phys.\ Rep.\ }
\def\prl{Phys.\ Rev.\ Lett.\ }

\def\prb{Phys.\ Rev.\ B\ }
\def\prd{Phys.\ Rev.\ D\ }

\def\jpa{J.\ Phys.\ A\ }

\def\jpsj{J.\ Phys.\ Soc.\ Jpn.\ }
\def\sci{Science\ }

\def\jsp{J.\ Stat.\ Phys.\ }
\def\cmp{Comm.\ Mat.\ Phys.\ }
\def\ijmpb{Int.\ J. Mod.\ Phys.\ B\ }

\def\ket#1{\left|{#1}\right>}
\def\bra#1{\left<{#1}\right|}
\def\braa#1{\left<{#1}\right.}

\begin{document}
\draft

\twocolumn[\hsize\textwidth\columnwidth\hsize\csname @twocolumnfalse\endcsname

\title{Short-ranged RVB physics, quantum dimer models and Ising gauge
theories}

\author{R. Moessner,$^1$ S. L. Sondhi$^1$ and Eduardo Fradkin$^2$}

\address{$^1$Department of Physics, Princeton University,
Princeton, NJ 08544, USA}

\address{$^2$Department of Physics, University of Illinois at Urbana-Champaign,
Urbana, IL 61801, USA}

\date{\today}

\maketitle

\begin{abstract}
Quantum dimer models are believed to capture the essential physics of
antiferromagnetic phases dominated by short-ranged valence bond 
configurations. We show that these models arise as particular limits of 
Ising ($Z_2$) gauge theories, but that in these limits the system
develops a larger local $U(1)$ invariance that has different consequences
on different lattices. Conversely, we note that the standard $Z_2$ gauge 
theory is a generalised quantum dimer model, in which the particular
relaxation of the hardcore constraint for the dimers breaks the
$U(1)$ down to $Z_2$. These mappings indicate that at least one realization
of the Senthil-Fisher proposal for fractionalization is exactly the
short ranged resonating valence bond (RVB) scenario of Anderson and of
Kivelson, Rokhsar and Sethna. They also suggest that other realizations
will require the identification of a local low energy, Ising link variable
{\it and} a natural constraint. We also discuss the notion of topological
order in $Z_2$ gauge theories and its connection to earlier ideas in 
RVB theory. We note that this notion is not central to the experiment
proposed by Senthil and Fisher to detect vortices in the conjectured
$Z_2$ gauge field.

\end{abstract}

\pacs{PACS numbers:
75.10.Jm, 
74.20.Mn 
}
]

\section{Introduction}

The question posed by high-temperature superconductivity is how a Mott
insulator becomes superconducting upon doping.\cite{pwabook} As the
insulator is itself, at low energies, also an antiferromagnet hostile
to the motion of holes, much work has been based on the notion that
the doped state is best approached from a ``nearby'' insulating state
that lacks long range order, \ie\ a spin
liquid.\cite{pwa87} The doped spin liquid is then argued to
become superconducting.\cite{fn-stripes}

The simplest such scenario casts the resonating valence bond (RVB) 
state proposed in 1973 
by Anderson\cite{Fazekas74} in the role of the spin
liquid. Pairs of electrons form singlet (valence) bonds, a superposition 
of which yields a liquidlike, non-Neel ground state. Holes doped into this 
state undergo spin-charge separation. The charge degrees of freedom, able 
to move freely through the spin-liquid, become superconducting upon Bose 
condensation. The spin excitations are understood as composites of 
spin-$1\over 2$ spinons and the decay of the electron into holon and spinon
provides a natural explanation of the broad quasiparticle spectra seen
over much of the normal state of the cuprates. 

RVB scenarios themselves cover a broad range of possibilities. The
short-ranged (SR) flavour of RVB stays close to Anderson's original
vision by including valence bonds only between electrons located in a
small neighbourhood of one another leads to gapped spinons
\cite{kivrokset}. Its low energy dynamics is believed to be captured
by the quantum dimer model (QDM) introduced in
Ref.~\onlinecite{Rokhsar88}, where a VB is represented by a dimer
linking the two electrons which form it.  Historically, the
short-ranged RVB was abandoned when the QDM failed to lead to a spin
liquid on the square lattice -- it typically leads to a columnar state
-- and was considered suspect for building in a spin gap (\ie\ a gap
to triplet excitations) that was not in evidence at optimal doping;
subsequent to the identification of the pseudogap regime and the
discovery of stripes these ``defects'' seem less compelling although
the problem of describing the collapse of the spin gap with doping is
still unsolved in this approach as is the still basic problem of
solving for the physics of a finite density of dopants.\cite{fn-subir}

In contrast the long-ranged RVB versions have received gauge theoretic
treatments based on $U(1)$ and $SU(2)$ reformulations of the Heisenberg
model that can give rise to a gapless mean-field spinon 
spectrum.\cite{ba} While the
broad similarity between the mean-field phase diagrams constructed
early on, and the phase diagram of generic cuprate superconductors
is striking, assessing the impact of fluctuations has been difficult.
In particular, the general belief that such gauge theories cannot give rise to
deconfined phases in 2+1 dimensions is at odds with the program of finding
a proximate fractionalized spin liquid.

Recently, Senthil and Fisher (SF),\cite{sentfish} building on earlier
work by Balents \etal,\cite{Balents98} have proposed to get around
this by reformulating the problem as an Ising gauge
theory.\cite{Wegner71} As Ising gauge theories {\it do} have
deconfined phases in 2+1 dimensions,\cite{kogut,savit} this seems
quite promising. What is not clear from their work, is exactly what
microscopic degrees of freedom are described by the Ising gauge
fields.\cite{fn-cd}
SF have offered two separate justifications for the presence
of Ising gauge fields. First, that a four fermion Heisenberg
interaction can be decoupled by means of an Ising gauge field and
second, that in models with separate electronic and superconducting
degrees of freedom, the latter can screen the charge of the former upto
a sign ambiguity in defining the needed square root of the cooper pair
operator. The former seems to us to be an interesting and exact
microscopic statement, but inconclusive regarding the nature of the
low energy theory; this point has also been made recently by 
Hastings\cite{mattgt}
and we will give a trivial example to illustrate this point later in
the paper. The second justification builds in the physics invoked in
earlier work, namely the capacity of a superconducting condensate to
screen charge and turn quasiparticles into spinons,\cite{rokkivspin}
 but it does appear
puzzling that it holds into insulating phases as hypothesized by SF.
In addition SF have argued that a deconfined phase involving Ising
gauge fields must be characterized by the notion of topological order
invoked in studies of the quantum Hall effect\cite{wenniu} 
and that this order
can be directly detected in an experiment.

In this paper we attempt to shed some light on these issues by showing
that at least one realization of Ising gauge physics is {\it exactly}
the physics of the short ranged RVB. We will do this by formulating
the QDM description of the latter as an odd 
Ising gauge theory, a term we will explain in 
Sect.~\ref{sect:qdmigt} below. In this we
will offer a variation on previous work by Fradkin and
Kivelson who mapped the problem onto a $U(1)$ gauge theory
instead.\cite{fradkiv}
This builds on a completely differently motivated stream of work of
two of us (with P. Chandra) on frustrated quantum Ising
models\cite{mcs2000} which has included a recent demonstration that
the quantum dimer model on a triangular lattice supports an RVB
phase.\cite{MStrirvb} This connection will allow us to interpret
various statements about Ising gauge theories in the language of
valence bonds -- it will turn out that the Ising variable {\it is} the
number of valence bonds -- and, we hope, make them easier to grasp and
evaluate. We should note that alternative identifications 
of spin liquid physics with Ising gauge theories in different limits 
have been made previously implicitly by Read and Chakraborty,\cite{readcha}
and explicitly by Read, Sachdev, Jalbert and 
Vojta,\cite{readsach1,sachjal,sachvojt} Wen,\cite{wen} 
and Mudry and Fradkin.\cite{mudry94}

A second benefit of this exploration is that it focusses attention on
what it takes to get an Ising gauge description of the low energy 
dynamics, namely a binary link variable and a local constraint. If
SF are correct and the Ising description has general applicablity,
it should be possible to make comparable identifications in other
contexts.

In the balance of the paper, we will review the QDM description of 
valence bond phases, describe the reformulation of the QDM as an 
Ising gauge theory and of general Ising gauge theories as generalised 
dimer models (GDMs), collate the known results on these models, discuss 
the notion of topological order in their context and conclude with 
a brief summary. As much of the interest of the paper lies in the
connection between QDMs and Ising gauge theories, we have felt it
useful to review a number of known results on both. It is useful
perhaps to list the results that are new to this paper. These
are the formulation of the QDM as an Ising gauge theory (Section
III), the introduction of the odd Ising gauge theory and its QDM limit 
(Section IV) and its identification with the action including a
Polyakov loop (PL) 
term written down by SF (Section IVB and Appendix 
B), the (elementary) treatment of Ising gauge theories in $d=1$ including
the presence of fractionalization in the odd case, the rationalisation 
of the QDM results on the square and triangular lattices via the absence
or presence of charge-2 Higgs fields (Section VB) and the discussion 
of topological order in Ising gauge theories (Section VI).

\section{Valence bond phases and quantum dimer models}

Consider an insulating magnet with enhanced quantum fluctuations (as
is the case with $S=1/2$ and other ``small'' values of the spin) and
competing interactions that frustrate any Neel ordering that would be
deduced from a semiclassical analysis. In the extreme case where any
residual order is extremely short ranged, there is a reasonable
expectation that the system will construct its ground state from
configurations in which all spins are paired in nearest neighbour
valence bonds. By continuity, we expect that there will be nearby
Hamiltonians for which valence bonds of finite length will suffice and
these are expected to share their basic physics with the purely
nearest neighbour case.\cite{chayes}
 As there is still a large number of short
ranged valence bond states possible, even with the restriction to this
sector there is a non-trivial problem remaining -- that of
diagonalizing the Hamiltonian within this highly degenerate manifold
-- which is the problem of ``resonance''.\cite{Fazekas74} 
Depending on the
details of the Hamiltonian, several phases might be realized. This set
of phases are what we call (short ranged) valence bond dominated
phases and, by hypothesis, they are all characterized by a spin gap.

There are two primary obstacles to investigating the physics of
valence bond dominated phases which we will restrict, in the
remaining, to those with purely nearest neighbor bonds. The first is
basic, namely the large degeneracy cited before, e.g. on the square
and triangular lattices in $d=2$ there are $e^{\alpha N}$\
($\alpha>0$) states on an $N$ site lattice.  The second is technical
in that different valence bond configurations are not orthogonal,
although their overlap is effectively exponentially small in the
number of dimers in which they differ. In some cases, there is also a
proof that they are linearly independent.\cite{chayes}

To deal with the second problem it is convenient to formulate an
expansion that can include the non-orthogonality perturbatively.  As
the parent configurations are in one-to-one correspondence with hard
core dimer coverings of the various lattices, such a tack leads to a
quantum dimer model. 

The sites the electrons reside on define the direct lattice.  The
Hilbert space of the QDM thus consists of all hardcore dimer coverings
of the direct lattice. The QDM Hamiltonian for the insulating case
(half-filling) consists of two parts, a kinetic ($\hat{T}$) and a
potential ($\hat{V}$) one. The former is off-diagonal and generates
the resonance plaquette moves between different dimer configurations,
whereas the latter, diagonal one, counts the number of plaquettes able
to participate in such resonance moves.

For the square lattice we find the Hamiltonian\cite{Rokhsar88}

\bea 
&&\lefteqn{H_{QDM} = -t \hat{T} + v \hat{V}=}
\nonumber \\ 
&\sum_\Box&\left\{-t\left
( |
\setlength{\unitlength}{3947sp}%
\begingroup\makeatletter\ifx\SetFigFont\undefined%
\gdef\SetFigFont#1#2#3#4#5{%
  \reset@font\fontsize{#1}{#2pt}%
  \fontfamily{#3}\fontseries{#4}\fontshape{#5}%
  \selectfont}%
\fi\endgroup%
\begin{picture}(155,154)(533,319)
\thicklines
\put(664,343){\circle{18}}
\put(556,342){\line( 1, 0){105}}
\put(557,343){\circle{18}}
\put(557,449){\circle{18}}
\put(556,449){\line( 1, 0){105}}
\put(664,449){\circle{18}}
\end{picture}
\rangle
\langle
\setlength{\unitlength}{3947sp}%
\begingroup\makeatletter\ifx\SetFigFont\undefined%
\gdef\SetFigFont#1#2#3#4#5{%
  \reset@font\fontsize{#1}{#2pt}%
  \fontfamily{#3}\fontseries{#4}\fontshape{#5}%
  \selectfont}%
\fi\endgroup%
\begin{picture}(154,155)(397,321)
\thicklines
\put(527,452){\circle{18}}
\put(528,344){\line( 0, 1){105}}
\put(527,345){\circle{18}}
\put(421,345){\circle{18}}
\put(421,344){\line( 0, 1){105}}
\put(421,452){\circle{18}}
\end{picture}
|+h.c.  \right) +v\left
( |
\setlength{\unitlength}{3947sp}%
\begingroup\makeatletter\ifx\SetFigFont\undefined%
\gdef\SetFigFont#1#2#3#4#5{%
  \reset@font\fontsize{#1}{#2pt}%
  \fontfamily{#3}\fontseries{#4}\fontshape{#5}%
  \selectfont}%
\fi\endgroup%
\begin{picture}(155,154)(533,319)
\thicklines
\put(664,343){\circle{18}}
\put(556,342){\line( 1, 0){105}}
\put(557,343){\circle{18}}
\put(557,449){\circle{18}}
\put(556,449){\line( 1, 0){105}}
\put(664,449){\circle{18}}
\end{picture}
\rangle \langle
\setlength{\unitlength}{3947sp}%
\begingroup\makeatletter\ifx\SetFigFont\undefined%
\gdef\SetFigFont#1#2#3#4#5{%
  \reset@font\fontsize{#1}{#2pt}%
  \fontfamily{#3}\fontseries{#4}\fontshape{#5}%
  \selectfont}%
\fi\endgroup%
\begin{picture}(155,154)(533,319)
\thicklines
\put(664,343){\circle{18}}
\put(556,342){\line( 1, 0){105}}
\put(557,343){\circle{18}}
\put(557,449){\circle{18}}
\put(556,449){\line( 1, 0){105}}
\put(664,449){\circle{18}}
\end{picture}
|+
|
\setlength{\unitlength}{3947sp}%
\begingroup\makeatletter\ifx\SetFigFont\undefined%
\gdef\SetFigFont#1#2#3#4#5{%
  \reset@font\fontsize{#1}{#2pt}%
  \fontfamily{#3}\fontseries{#4}\fontshape{#5}%
  \selectfont}%
\fi\endgroup%
\begin{picture}(154,155)(397,321)
\thicklines
\put(527,452){\circle{18}}
\put(528,344){\line( 0, 1){105}}
\put(527,345){\circle{18}}
\put(421,345){\circle{18}}
\put(421,344){\line( 0, 1){105}}
\put(421,452){\circle{18}}
\end{picture}
\rangle \langle
\setlength{\unitlength}{3947sp}%
\begingroup\makeatletter\ifx\SetFigFont\undefined%
\gdef\SetFigFont#1#2#3#4#5{%
  \reset@font\fontsize{#1}{#2pt}%
  \fontfamily{#3}\fontseries{#4}\fontshape{#5}%
  \selectfont}%
\fi\endgroup%
\begin{picture}(154,155)(397,321)
\thicklines
\put(527,452){\circle{18}}
\put(528,344){\line( 0, 1){105}}
\put(527,345){\circle{18}}
\put(421,345){\circle{18}}
\put(421,344){\line( 0, 1){105}}
\put(421,452){\circle{18}}
\end{picture}
|
\right)\right\}\, ,
\nonumber \\
&&
\label{eq:exqdm}
\eea
where we have kept only the simplest kinetic and potential
energy terms with coefficients $t$ and $v$, and the sum $\sum_\Box$\
runs over all plaquettes. In what follows, 
we refer to the QDM with $v\neq0$\ as the extended QDM, whereas QDM 
on its own refers to the case $v=0$. On other lattices $\hat T$
will take the form of a sum of resonance moves on the shortest even
loop (which is a plaquette in this case) and $\hat V$ 
will be count the number of such possible moves in a given
configurations, e.g. on the honeycomb lattice both will involve
three dimers.

There are two main paths for obtaining a quantum dimer model from a
magnetic system. The first is based directly on an $SU(2)$ Heisenberg 
magnet, the second
on ``large-N'' generalisations thereof. The former\cite{Rokhsar88}
uses the abovementioned small overlap to generate a perturbation
expansion.
The valence bond states can be labeled by orthogonalised dimer
configurations, the quantum dynamics of which is captured by the
leading order dimer plaquette resonance move, $\hat{T}$.  The leading
order diagonal term is given by $\hat{V}$.  The second
path\cite{sachlect} generalises the $SU(2)\sim Sp(1)$ to $SU(N)$
on bipartite lattices or $Sp(N)$ generally. The latter method
essentially generalizes the Schwinger boson representation for
the spins  by introducing a large number of additional boson flavours.
In the limit $N\rightarrow\infty$, taken at a fixed number of
bosons per site, the ground states at leading order can be labelled by
dimer configurations. It is the next order, $1/N$, terms which
generate the abovementioned dimer resonance move. 

\section{The quantum dimer model as an Ising gauge theory}
\label{sect:qdmigt}

\begin{figure}
\epsfxsize=2.5in
\centerline{\epsffile{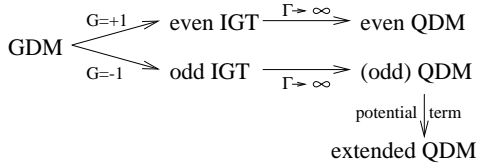}}
\vskip5mm
\caption{Naming conventions used in this paper. Theories without
matter are referred to as pure, with matter as doped. The even IGT is
dual to a ferromagnetic transverse field Ising model, the odd IGT to a
fully frustrated one (see Appendix~\ref{app:dual}), with the
$\Gamma \rightarrow \infty$\  limit corresponding to a projection onto 
the magnetic ground state(s) of the dual Ising models.}
\label{fig:convention}
\end{figure}

In the following, we discuss the relationships between a number of
different models of interest in the context of high-temperature
superconductivity and quantum magnetism. The naming conventions for
them are depicted in Fig.~\ref{fig:convention}.  Our first mapping --
of the QDM to {\it an} Ising gauge theory (IGT) -- proceeds as
follows. The naive Hilbert space (inclusive of gauge equivalent
states) of any IGT is defined by an Ising variable $\sigma=\pm 1$\ on
each link of the lattice; each variable will be taken to be the
eigenvalue of an operator $\hs^x$\ on the corresponding link. We can
identify the link variable with the presence or absence of a dimer on
the link, \ie\ the number of dimers on each link is now given by
$n=(1+\sigma)/2$ and the dimer number operators are $\hat{n} =
\frac{1}{2}(1+\hs^x)$, where we have suppressed the link index.

Evidently, the naive Hilbert space is too big and we must identify the
physical subspace that corresponds to allowed hardcore dimer
coverings, which is done by imposing a constraint at every site that
only one link emanating from it be occupied by a dimer.  This is
expressed as an operation, $\hat{G}$, which leaves invariant only the
physical states, $\phy$, those fulfilling the hardcore condition.
In terms of the operators $\hs^x$, the hardcore constraint
becomes: \bea \sum_{+} \hs^x\phy=(-n_c+2) \phy
,
\eea
where the sum is over all the links emanating from a given site $i$,
and $n_c$\ is the coordination of that site. This implies
\beq
\hat{G}_\alpha\phy=\phy
\eeq where
\beq
\hat{G}_\alpha\equiv
\exp \left(i \alpha\sum_+(\hs^x+1-2/n_c)\right)
\eeq
for {\em any} $\alpha$ at {\em each} site.

To write the Hamiltonian at half filling, $\hat{H}_I$, in a form such
that $[\hat{G},\hat{H}_I]=0$, we define the usual spin-half raising and
lowering operators $\hs^{\pm}=\hs^y\pm i\hs^z$, which
respectively add and remove a dimer on a link. We obtain: \bea
\hat{H}_I&=&-t\hat{T}+v\hat{V} =-t\sum_{\Box}\left(
\hs^+\hs^- \hs^+\hs^- +h.c.\right)
\label{eq:hi}
\\ &&+\frac{v}{4}
\sum_{\Box}
\left( (1+\hs^x_1)(1+\hs^x_3)+(1+\hs^x_2)(1+\hs^x_4)\right)\ ,
\nonumber
\eea
where the sites in the last term are labelled sequentially 
around the plaquette, $\Box$. The generalization to other
lattices follows the prescription for writing down QDMs outlined
earlier.

Invariance of $\hat{H}_I$ under local Ising gauge operations is easily
checked. In fact a larger, $U(1)$, symmetry
arises because $\hat{G}_\alpha^{-1}\hs^\pm\hat{G}_\alpha=\exp(\pm
i\alpha) \hs^\pm$ so that the phases picked up by the products in
$\hat{T}$ cancel.  As we will discuss further below, this local
$U(1)$\ is a consequence of local dimer number conservation.  So we 
have the situation that while the Hilbert space is that of an Ising 
gauge theory, the physical states and Hamiltonian are invariant under 
a set of continuous local gauge transformations that have the form of a
$U(1)$ gauge theory. This is a local gauge theory version of the 
more familiar quantum $S=1/2$\ XY-model, which also lives in an 
Ising Hilbert space.

When holes are doped into the valence bond manifold, we need
to worry about the comparison of the (potentially) large hole
hopping energy $t$ to the existing magnetic scales that are
responsible for the triplet gap. In the spirit of RVB theory,
we assume that one can approach the problem from small values
of $t$ and that it is therefore sufficient to keep states
in which the spins pair up into valence bonds except at the
sites from which they have been removed. The effect of the
hole kinetic energy is to move a hole and a dimer in ways
that keep within this manifold.\cite{fradkiv}

In the Ising gauge language, we have to add a matter (Higgs) part,
$\hat{H}_m$, to the Hamiltonian. Let the Ising Higgs field be denoted
by an operator $\htau_x$\, its eigenvalues $\pm 1$\  denoting the 
presence/absence of a hole. Since each site has
either a dimer ending on it or is occupied by a hole, the constraint
equation is modified to
\bea
\hat{G}\phy &=&
\exp{\left(i \alpha(\htau_x+
\sum_\Box(\hs^x+1-1/n_c))\right)}\phy
\nonumber \\
&=&\phy\, ,
\eea
and the full Hamiltonian is given by $\hat{H}=\hat{H}_I+\hat{H}_m$,
with
\bea
\hat{H}_m=-m \sum_{(ijk)}\left(
\htau^-_i\hs^+_{ij}\hs^-_{jk}\htau^+_k+h.c.
\right)\, ,
\label{eq:Hm}
\eea
and the sum on $(ijk)$\ running over triplets of neighbouring sites.

\section{Ising gauge theories as generalized dimer models}

We now review the converse logic of starting with the more familiar
action of the standard IGT (referred to as even IGT from hereon) and
ending up with a Hamiltonian formulation. The resulting theory is a
generalized dimer model (GDM) which features dimers on links that obey
a generalized dimer constraint -- in the even case this requires an
even number of dimers to emanate from each site. We also show
that the addition of the 
Polyakov loop term $S_{P}$\
to the standard action introduced by SF leads to
a Hamiltonian formulation that is a GDM with the constraint of an odd
number of dimers per site and that in a special limit, it reduces to
precisely a quantum dimer model with the hard core dimer constraint.

\subsection{Hamiltonian vs.\ Lagrangian formulation}

Consider the action of the standard IGT, hereafter referred to as 
the even IGT,
\beq
S=-K^\tau\sum_{\Box}\hsi^z\hsi^z\hsi^z\hsi^z
-K^s\sum_{\Box}\hsi^z\hsi^z\hsi^z\hsi^z \ .
\label{eq:SIGT}
\eeq made anisotropic by choosing a coupling, $K^\tau$, for plaquettes
containing links in the imaginary time (temporal) direction, different
from that for purely spatial plaquettes, $K^s$.\cite{kogut} This is
neccessary to take the time continuum limit needed in the derivation
of the Hamiltonian.

We can now choose a gauge wherein all $\hsi^z$\ in the time direction
are +1, so that the first term in Eq.~\ref{eq:SIGT}
becomes a simple bilinear, $-K^\tau\sum_\Box\hsi^z\hsi^z$, involving
only the links in the space directions. (Strictly speaking, it is
not possible to do this as it would have the effect of modifying the
gauge invariant products of $\sigma^z$ along temporal loops. However
this obstruction is not important in the time continuum limit in the
case of the even IGT at $T=0$.)
One then establishes the
equivalence to an appropriate Hamiltonian {\it and set of constraints}
in one dimension less by comparing the expressions for the partition
function generated by this action to that arising from a
Trotter-Suzuki decomposed path integral formulation generated by the
Hamiltonian \beq \hat{H}_{GDM}=
\Gamma\sum_{-}\hs^x-\kappa\sum_\Box\hs^z\hs^z\hs^z\hs^z \ ,
\label{eq:HIGT}
\eeq 
where the first sum runs over all links and the Hilbert space
is limited by constraints (below). One finds an
equivalence of the partition functions in the limit
$K^\tau\rightarrow\infty$, with $K^s=\kappa \exp(-2K^\tau)$\ fixed.

The Hamiltonian defined in Eq.~\ref{eq:HIGT} retains a gauge
invariance under flipping all spins (in the $\hsi^z$\ basis) emanating
from one site. This transformation is generated by $\hat{G}_{\rm
IGT}=\prod_+\hs^x$. To reproduce the physics of the even IGT we need
to impose $\hat{G}_{\rm IGT}= 1$ at every site.

While the Hamiltonian and Hilbert space are naturally derived
in the $\hsi^z$\ basis, the meaning of these constraints becomes
transparent by considering the system in the $\hsi^x$\ basis. For
$\hat{G}_{\rm IGT}(i)\phy=+\phy$, an even number of links emanating
from site $i$\ has $\sigma^x=1$. Identifying, as
we did in Sect.~\ref{sect:qdmigt}, the presence (absence) of a dimer
with $\sigma^x=\pm1$, we see that the constraint $\hat{G}_{IGT}=
+\hat1$\
implies the presence of an {\it even} number of dimers emanating from
each site -- whence our label ``even'' for the IGT under consideration.

\begin{figure}
\epsfxsize=3in
\centerline{\epsffile{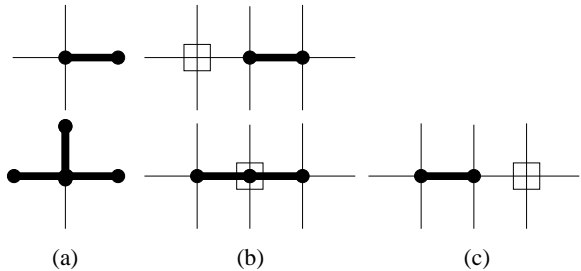}}
\vskip 0.5cm
\caption{ (a) Top: Allowed hardcore dimer configuration. Bottom:
Allowed configuration in an odd IGT but not QDM. (b) Hopping process
of hole (denoted by square) generated by $\hat{H}_u$, which does not
conserve dimer number.  Two applications of $\hat{H}_u$\ yield an
allowed final QDM configuration (c); the net hop is generated directly
by $\hat{H}_m$. }
\label{fig:hilbert}
\end{figure}

\subsection{The odd IGT and the Polyakov loop term $S_{P}$}

To obtain a Hamiltonian problem in which the physical states have an
{\it odd} number of dimers at each site (the odd IGT), we need to add
the Polyakov loop term to the action above:
\bea
e^{S_{P}}=\prod_t \hsi_t^z\ ,
\eea
where the product runs over all temporal links.
 This is equivalent to assigning a {\it sign} to
each space-time configuration which is the product of
Polyakov loops\cite{polloop}
 in the temporal direction that wrap
around the system for each spatial site. It can
be shown (see Appendix~\ref{app:berry}) that this
is equivalent, in the time continuum limit, to choosing $\hat{G}_{IGT}=
-\hat1$\ in picking physical states for the action of the 
Hamiltonian Eq.~\ref{eq:HIGT}. We should note, that even with isotropic
couplings in the action Eq.~\ref{eq:SIGT}, 
$S_{P}$ breaks the symmetry between
space and time (lattice Euclidean invariance). Consequently one
may need to be careful about distinguishing the behavior of Wilson
loops in space and those in time (Polyakov loops) in distinguishing
confined and deconfined phases -- the latter are then the correct quantity
to calculate.

For the square lattice, the inclusion of $S_{P}$ (which arises in 
the work by SF for a Mott insulator with an odd number of electrons per 
site) thus represents a mixture of dimers (one link occupied) and tetramers
(three).\cite{fn-tetra} Whereas the dimers are amenable to an obvious
physical interpretation as valence bonds, we are not aware of any
similar interpretation of more complicated polymers.

We thus see that the somewhat unconventional form of the kinetic term
in Eq.~\ref{eq:hi}, which consists of raising and lowering operators,
rather than simply $\prod\hs^z$, arises from the desire to preserve
the hardcore dimer constraint manifestly.

\subsection{QDM limit of odd IGT}

One can nonetheless retrieve the hardcore constraint by
explicitly removing the supernumerary dimers `by hand', through a very
large coupling constant $\Gamma$. In this limit (the
$\Gamma \rightarrow \infty$\
limit), the term $\prod_\Box\hs^z$\ becomes equivalent
to the kinetic term of Eq.~\ref{eq:hi}; it is in this limit
that the IGT of Senthil and Fisher including the $S_{P}$\ term is
equivalent to our QDM. In this context, it is interesting to note that
the original $U(1)$\ gauge theory of Fradkin and
Kivelson\cite{fradkiv} is close in spirit to the above
construction. There, the presence of a dimer is encoded by an angular
momentum variable on each link, $L_{ij}$, which is restricted by an
analog of the transverse field term to values 0 or 1. The raising and
lowering operators (conjugate to the $\hat{L}_{ij}$) appear in the
kinetic term $\hat{T}$.

In the presence of doping, there is an additional difference between
the two theories which lies in the nature of the allowed hopping
terms. The GDM admits terms of the form 
\bea
\hat{H}_u+\hat{H}_\lambda=-u\sum_{-}
(\htau^+_i\hs^z_{ij}\htau^-_j+h.c.)+\lambda\sum_.\tau_i^x,
\eea
corresponding to the processes depicted
in Fig.~\ref{fig:hilbert}b. (The sum $\sum_.$\ runs over all sites
of the lattice).
Note that a notion of charge conservation and hence a global
$U(1)$ invariance,
mandates the use of $\htau^\pm$\ operators rather than $\htau^z$.  
In the  $\Gamma\rightarrow \infty$\ limit, this term becomes 
ineffective since one of the two configurations which $\hat{H}_u$\ 
connects is projected out. However, for $\Gamma$ large but not infinite
we generate the term $\hat{H}_m$\ (Eq.~\ref{eq:Hm}) with $m \sim
{u^2/ \Gamma}$ to obtain dynamic holes (see Fig.~\ref{fig:hilbert}).

Quite generally, extended QDMs thus arise as liimits of odd IGTs with
additional couplings.
>From the perspective of IGTs, this is a simplification which focuses
attention on the existence of larger local $U(1)$ invariance, but not 
much more. However from the perspective of the physics of 
antiferromagnets the QDM along with its associated conservation of
the number of dimers (valence bonds) per site appears to be the
more natural construction. It also has the appeal that it allows
two question basic to setting up a gauge theory to be answered
transparently: 1) what is the link variable? and 2) what is the
local constraint? As we have noted, in the QDM limit this leads
to an Ising link variable which is simply the number of valence bonds
but a constraint on their number more appropriate to a $U(1)$ theory. 
At the very least, short ranged RVB theory is an example of an
IGT of a strongly correlated system and we will use it to examine
some of the observations made by SF about IGTs in general. More
generally, it seems to us that in order to make the case for an
IGT description in other phases, it would be extremely useful to
have a comparable identification of the link variable and the
constraint (even in some appropriate coarse grained sense).

\subsection{Duality to Ising models in $d=2+1$}

It is well known that IGTs in $d=3$\ are dual to Ising
models.\cite{kogut,savit} In the Hamiltonian formulation in $d=2+1$,
there is a simple, geometrical, way of seeing this which
demonstrates how the standard constraint $G=+1$\ translates into a
dual ferromagnetic transverse field Ising model while the alternative 
constraint $G=-1$\ translates into a fully frustrated Ising model
in a transverse field. An account of this mapping is given in
Appendix~\ref{app:dual}.

Evidently, it is most economical to study these dual Ising models.
An extensive study of frustrated Ising models in $d=2+1$ is reported
in Ref.~\onlinecite{mcs2000}.

\section{Phase Structure and Quasiparticle Fractionalization}

For all the pure (undoped) QDMs, the question of primary interest is 
whether
they possess a dimer liquid or RVB phase. Such a phase automatically
leads to free spin-1/2 excitations (spinons) and to the decay of an
ejected electron into a spinon and a spinless charged hole (holon)
which provides an example of spin-charge separation (in an insulator)
in general dimensions.

The physical arguments leading to the above conclusions are simple
in the valence bond language.
A valence bond can be broken up into two neighboring spins 1/2.
In a valence bond liquid the cost of separating these two objects
to infinite separation will be finite -- hence the existence of
a spinon spectrum above the triplet gap. Further, at large separations
one can remove one of the spinons to obtain a spinon-holon pair
that has the quantum numbers of a missing electron, or hole. Hence
in a photoemission experiment one will see a fractionalized 
spectrum above the charge gap.

Within the framework of the dimer models both spinons and holons
are represented by monomers and the issue is one of computing
the free energy of the system as a function of monomer seperation.
In a liquid phase this will be finite.

It is worth digressing a bit and noting the translation between the
standard gauge theory lore and the above statements. The standard
diagnostic of confinement in a pure gauge theory is the Wilson
loop. In a Lorentz invariant theory its orientation does not matter
and hence we may compute the expectation value of a spatial loop as
well as a temporal loop which is directly related to the energy of two
separated quarks. In dimer models one does not have Lorentz
invariance -- trivially for we are in the time, but not space,
continuum limit and less trivially because the Polyakov loop
term is not Lorentz invariant, as already noted.
Consequently one should really compute the temporal Wilson loop 
(the Polyakov loop). 
Nevertheless, as the following calculation shows, liquidity implies a 
perimeter law for the spatial Wilson loop, and 
as explained above, 
liquidity in the dimer models signals deconfinement.

\begin{figure}
\epsfxsize=3in
\centerline{\epsffile{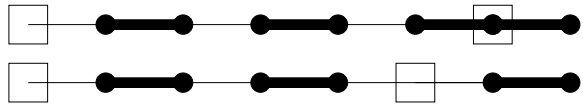}}
\vskip 0.5cm
\caption{ Top: Neutral `quark-antiquark' pair. Bottom: Charge-2 pair.
A square denotes a hole.}
\label{fig:quark}
\end{figure}

A second caveat is that in the dimer model
it is the two monomer free energy that has a clear meaning. As this is
a state with physical charge 2 relative to the ground state, its free
energy cannot be computed as a neutral vacuum correlator (unlike a
quark/anti-quark potential). The strict analog of the
quark/anti-quark potential is the interaction between a monomer and a
site with two valence bonds (Fig.~\ref{fig:quark}). 
Presumably the long distance
interactions in the two cases will track one another.

\begin{figure}
\epsfxsize=3in
\centerline{\epsffile{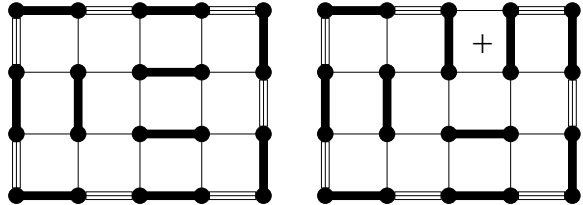}}
\vskip 0.5cm
\caption{Evaluation of a Wilson loop, taken around the circumference
of the displayed region. The string of operators $\pi_W$\
annihilates the state unless it encounters an alternating sequence of
occupied and empty bonds (left panel). The links occupied by dimers
after the action of $\pi_W$\ are denoted by empty rectangles.
Flipping only one plaquette (marked by a plus, right panel) leads to
the configuration being annihilated by $\pi_W$.  }
\label{fig:wilson}
\end{figure}

With these comments, consider the spatial Wilson loop in the
dimer limit. The product
\bea 
\pi_W\equiv\prod_{i=1}^{L_c}\hsi^z
\eea
reduces to the strings
\bea
\pi_W=(\hsi^+_1\hsi^-_2 ... \hsi^+_{L_c-1}\hsi^-_{L_c}+h.c.)
\label{eq:string}
\eea 
of dimer creation and destruction operators. In taking the
expection value of $\pi_W$, we select pieces of the ground state
wavefunction that contain precisely the dimer strings in
Fig.~\ref{fig:wilson} along the selected loop. To estimate this
fraction we appeal to the extensive entropy of dimer configurations
and to a healing length $\xi$ in a dimer liquid. In the liquid we
therefore obtain an estimate \bea \pi_W \sim e^{-c(\xi)L_c} \ , \eea
that exhibits a perimeter law consistent with the lore for a
deconfined phase.\cite{kogut}
 Here, $c(\xi)$\ is some numerical constant depending
on the correlation length.  Strictly speaking we have carried out an
estimate for a ground state wavefunction spread equally over all dimer
configurations -- such states arise at the so-called Rokhsar-Kivelson
points (see below) -- but this will be qualitiatively correct
everywhere in a liquid phase.

We should note that the above considerations are for static matter
interacting via fluctuating gauge fields or dimers. While it
is entirely reasonable that a deconfined phase at zero doping will
continue into a deconfined phase at finite doping, the finite
doping problem is logically distinct and needs to be treated
carefully on its own. We note, for instance, the well-known result
that dynamic matter does not permit an area law phase for the 
Wilson loop, even at arbitrarily weak coupling on account of
the screening of the gauge force.\cite{fradshen}

In the remainder of this section, we discuss the properties of dimer
models on a number of lattices and for varying dimensionality. Besides
presenting a number of new results, especially in $d=1$, we collate 
several results from the literature, several of which are 
in the guise of stacked magnets or transverse field Ising
magnets, which therefore need translating. We will also refer to
results on the GDMs or IGTs outside the QDM limit in places.

\subsection{$d=1$}

While $d=1$ is special, it is instructive in that it {\it does} provide
an example of fractionalization that though distinct from the
higher dimensional versions, fits nicely into the QDM description.
This point was overlooked by SF in their analysis of IGTs.

Consider first the pure even IGT. In this case there are only
two states in the Hilbert space, those with $\sigma^x=1$ for
$\sigma^x=-1$ on all links. The Hamiltonian can only count the number
of dimers, as there is no local resonance move:
\bea 
\hat{H} = \Gamma \sum_- \hs^x\ ,
\eea whence the $\sigma^x=-1$ state is
always the ground state. Consider introducing two holes at a
separation $R$. The constraint now requires that the links between the
holes carry $\sigma^x=1$ which leads to an energy cost linear in $R$
and hence confinement. This is the well-known result on the purely
confining character of the $d=1$ (even) IGT.

Interestingly, the odd IGT behaves very differently! There are still
two states in its Hilbert space, but they consist of states with
alternating values of $\sigma^x=-1$ and $\sigma^x=1$, \ie\ dimers and
no dimers -- evidently there are two such states related by a
translation. These states are degenerate in energy. Consequently the
introduction of a hole still produces a domain wall between the two
phases, but two domain walls do not attract -- the charge carriers are
deconfined solitons. This is, of course, the familiar mechanism of
solitonic fractionalization from studies of conducting 
polymers.\cite{polymerfract}  
What
is interesting is that the odd IGT captures this mechanism in $d=1$
automatically.

This is a good place to give a trivial example of the
difference between microscopic rewritings and effective gauge theories.
Consider the one-dimensional Heisenberg chain with first $J_1$ and 
second neighbor $J_2$ antiferromagnetic interactions. This can be
rewritten as an IGT with two Ising gauge fields -- one for each bond.
As is well known, for sufficiently large $J_2$ the chain is in the 
dimerized (Majumdar-Ghosh) phase where the effective theory clearly
involves just one Ising gauge field. 

\subsection{$d=2$}

Two dimensions is the case of maximum interest in the context of theories
of cuprate superconductivity. 

\subsub{Square lattice} A general analysis of the even IGT coupled to
Ising matter has been given a while back by Fradkin and
Shenker.\cite{fradshen} They showed that two phases exist: a
deconfined phase with free charges in the spectrum and a
confined/Higgs phase. The phase diagram of the odd theory is not known
in as much detail. What is known is that the undoped odd IGT has a
confinement transition accompanied by translational symmetry breaking
as the QDM limit is approached. This result follows
from analyses of the dual transverse field Ising model\cite{mcs2000}
as well as from a map from the QDM to a height model.\cite{henleyjsp}
Consequently, the purely kinetic QDM on the square lattice gives rise
to a valence bond crystal with confined spinons.

The extended QDM (Eq.~\ref{eq:exqdm}) has been studied in detail in
and found to be ordered for all values
of $v$, except for a transition at the Rokhsar-Kivelson (RK) point $v=t$\
between a staggered ($v>t$) and a four-fold degenerate state ($v<t$) 
which is likely a plaquette state close to the RK point and then
gives way by a first order transition to a columnar state at
large negative $v$.
\cite{Rokhsar88,fradkiv,subirfinitesize,levitov,readsach3,leung}
At the RK point, 
the ground state is an equal amplitude superposition of all dimer 
configurations.
Spinons are
deconfined precisely at the transition only, and confined
elsewhere.\cite{MStrirvb} The unusual feature that a critical point
intervenes between two crystals, finds an elegant explanation in terms
of height representations: the effective action in the proximity of
the RK point has the form conjectured most completely by 
Henley\cite{henleyjsp}
\bea
S \sim \int d^2 x\ d\tau\, \left[(\partial_\tau h)^2 + \rho_2 (\nabla h)^2 +
\rho_4 (\nabla^2 h)^2 \right]
\eea
with $-\rho_2 \propto (v/t) + 1$ changing sign precisely at the 
RK point. This action accounts for the crystal for $v<t$ which is a 
flat state of the height variable, the critical correlations and 
resonon spectrum $\omega \sim k^2$
at the RK point, and the staggered state for $v>t$ which corresponds
to the maximum tilt of the height variable. It also accounts for two
other non-trivial features of the RK point, namely that it has 
degenerate ground states in all winding number sectors (the RK point
action is insensitive to gradients of heights) and that its equal
time height correlations are logarithmic and precisely those of the 
classical dimer problem -- which follows from the observation,
$$\int d\omega {1 \over \omega^2 + \rho_4 k^4} \sim {1 \over 
\sqrt{\rho_4} k^2} \ .$$

Less is known about the doped QDM, or the odd IGT coupled to charged
matter. A plausible scenario, based on a different large $N$ limit
than the one that gives rise to the QDM, has been advanced by Sachdev
and Vojta\cite{sachvojt} who find a set of striped states followed by
a d-wave superconductor. A direct treatment of the doped QDM has not
been carried out except for an early mean-field theory by Fradkin and
Kivelson,\cite{fradkiv} which led to an s-wave superconducting saddle
point, but which treats electron hopping terms which lack crucial
phase factors stemming from the microscopics. Further work on the
doped QDM would certainly be desirable.

\subsub{Triangular lattice} The standard IGT on the triangular lattice
(defined by taking products of $\sigma^z$ around elementary triangles)
is dual to the ferromagnetic transverse field Ising model on the
honeycomb lattice.\cite{savit} Consequently, it has a phase transition
identical to that of the square lattice IGT, between a confined and
deconfined phase. While we do not know of a detailed analysis of the
problem with Ising matter coupled to the Ising gauge field, we expect
that the Fradkin-Shenker analysis applies. 

The odd IGT  is dual to the fully frustrated Ising
model on the honeycomb lattice (see Appendix~\ref{app:dual}),
which has been studied by Chandra and
ouselves\cite{mcs2000} with some evidence for weak ordering involving
the breaking of translational symmetry in a confining phase. It is
also possible that the confining phase is absent altogther. Indeed, in
the extended QDM we have shown that there is definitely a liquid phase
for a finite range of parameters, $2/3\alt v/t\leq1$, in which spinons
are deconfined.\cite{MStrirvb} This is the only known example of a
deconfined phase in an IGT in the QDM limit. We
note that a recent neutron scattering experiment on a triangular
magnet, albeit a spatially anisotropic one, appears to have detected
deconfined spinons.\cite{coldea}

\subsub{Other lattices} The results for the QDM can be generalised to
other lattices.  The behaviour of quantum dimer models on bipartite
lattices follows that of the square lattice. This is a consequence of
the equivalence of the classical dimer models to a height
model,\cite{Blote82,henleyjsp} which in $d=2$\ implies critical
correlations which result in an ordering transition when quantum
fluctuations are switched on.\cite{mcs2000} In particular, this class
includes the hexagonal lattice quantum dimer model.\cite{fn-baroque}

For QDMs on non-bipartite models, no such general result is
known. Typically, one expects QDMs to be disordered and gapped at the
RK point $v=t$, and we suspect that this will result in an extended
disordered phase for $v\alt t$ with spin charge separation in analogy
to the triangular lattice case.
For some interesting results on depleted lattices see 
Ref.~\onlinecite{shtengel}.

\subsub{Charge-$2$ Higgs scalars}
The difference between the results for bipartite lattice and the
triangular lattice can be rationalized by a mechanism that has been
invoked several times in previous work on the subject. In the QDM
limit, we obtain a theory with a $U(1)$ gauge symmetry as we noted
earlier, albeit one that lives in an IGT Hilbert space. 

A celebrated result of Polyakov\cite{poly77} showed that the pure
compact $U(1)$ gauge theory confines at all values of its coupling in
2+1 dimensions.  This theory, defined by the Maxwell Lagrangian in the
continuum, is naturally formulated on the bipartite square lattice --
the plaquette product naturally translates into the former
Lagrangian. It was argued by Fradkin and Kivelson\cite{fradkiv} that
the Polyakov argument goes through for the IGT with the $S_P$\
term. It follows then that we should expect the QDM limit of the odd
IGT to be confining. Evidently, such an argument does not by itself
rule out a deconfined phase in the extended QDM and one needs the more
detailed height representation analysis to show that.

A second important result on $U(1)$ gauge theories is due to Fradkin
and Shenker who studied their phase structure when coupled to
matter.\cite{fradshen} They showed that a coupling to charge-1 matter
fields does not allow a deconfined phase to exist but that a coupling
to charge-2 scalars {\it did} allow a deconfined phase to exist. This
result provides a rationalization of the triangular lattice results:
to make contact with the standard results on the square lattice, one
must treat the triangular lattice as a square lattice with additional
diagonal bonds.  These additional bonds imply
the presence of a charge-2 scalar field coupled to the gauge field,
which then opens up the possibility of a deconfined phase.

The general attractiveness of the Fradkin-Shenker result to workers on
two dimensional quantum magnetism is evident. Heisenberg models are
easily reformulated as $U(1)$\ gauge theories but in the search for
spin liquids with deconfined spinons this is an
embarassment. Consequently it has been suggested by many workers:
Read and Sachdev,\cite{readsach1} as well as Mudry and 
Fradkin\cite{mudry94}) that the condensation of an
appropriate charge-2 scalar field would allow a spin liquid to
exist. In his work on the $Sp(N)$ analysis of the triangular
lattice,\cite{sachtri} which exhibits a disordered phase, Sachdev
again argued that fluctuations about the saddle point solution had the
structure of a $U(1)$ gauge field coupled to a charge-2 scalar, for
essentially the same reason we invoked above.

\subsection{$\lowercase{d}>2$}
In $d>2$, little is known about the properties of dimer models,
even in the classical case. Formally, the Pfaffian methods used in
$d=2$\ to gain information about the classical models break down due
to the overwhelming increase in the number of terms to 
be evaluated in higher dimensions.\cite{zecch}

Within the framework of dimer models, it is likely that spin-charge
separation becomes prevalent because dimer models should become more
disordered in high dimensions. This can be rationalised as the
hardcore constraint cannot be spread out but number of possible
orientations increases. 

However, the usefulness of quantum dimer models for describing the
physics of Mott insulators/Heisenberg models decreases in higher
dimensions as it becomes increasingly hard energetically to stabilise
valence-bond dominated configurations against the Neel state.

Senthil and Fisher have proposed an experiment to test their ideas on
a fractionalized phase in the cuprate phase diagram and have
explicitly linked this experiment to the notion of topological order
invoked previously in studies of the fractional quantum Hall effect by
Wen and co-workers.\cite{wenniu} We examine these assertions in the
context of dimer models, in reverse order. In particular we will
be interested in the connection between their ideas and the topological
analysis of valence bond 
states.\cite{thoulessquant,Rokhsar88,bonesteel,MStrirvb}

\subsection{Topological order in the quantum Hall 
effect} 
Quantum Hall states do not break any
obvious symmetry captured by a local operator; this excludes
cases, such as quantum Hall ferromagnets, where a symmetry
may be broken in addition. There are two alternative approaches
to characterizing quantum Hall states which can both be derived from
the rewriting of the electron dynamics in terms of bosons
coupled to one or more fluctuating Chern-Simons gauge 
fields.\cite{zhk,nr,gm} 
In the first, one focuses on the bosons and characterizes their 
condensation via an infinite particle electron operator which 
works everywhere in the quantum Hall phase that grows out of the
ideal quantum Hall state in a disordered system.\cite{sondhigelfand}
In the second approach, one integrates out the bosons to obtain a 
purely gauge action, which then contains the Chern-Simons term as its 
leading piece.

The Chern-Simons term is topological, \ie\ it is insensitive to
the metric of the manifold it is defined on. The pure Chern-Simons theory,
which describes the strict infrared behavior of the quantum Hall system,
has a finite dimensional Hilbert space with a set of degenerate
states whose number depends on the topology of the manifold.\cite{witten}

This leads to the notion of topological order -- the idea that a
quantum Hall state can be characterized by its ``response'' to the
topology of the underlying manifold. Operationally, one imagines
computing the exact spectrum in finite volumes and looking for a
low-lying cluster of states clearly (\ie\ parametrically in system
size) separated from all other states. This works perfectly for
clean quantum Hall systems on the torus -- e.g. there are $q$
exactly degenerate states at filling factor $1/q$. In this case
it is also possible to contruct operators, corresponding to the
adiabatic insertion of one quantum of flux through the holes
that have the effect to transforming one ground state into another.
As these operators commute with the Hamiltonian, their failure
to leave the ground state invariant was interpreted by Wen and
Niu as the breaking of a topological symmetry (the symmetry
algebra itself being dependent on the topology of the manifold).

To summarize: Topological order in clean quantum Hall systems
at the ideal filling factors involves,\\
a) a ground state multiplet, separated from other states by
an amount parametrically larger than the splitting between
them, and with a degeneracy that increases with the genus, $g$,
of the manifold as $q^g$\\
b) a topological symmetry algebra containing operators that
move the system between different members of the ground state
multiplet \\
c) a long wavelength action (the Chern-Simons action) that defines
a theory with a finite dimensional Hilbert space with the same
degeneracies.

In the clean system, quantum Hall states compete with the Wigner
crystal, or with various charge density wave states. The latter
pair of states will lead to a higher ground state degeneracy,
indeed an infinite degeneracy in the infinite volume limit 
corresponding to the various translations of the crystals as a
whole. The same is true for the quasiparticle Wigner
crystals that will form in the close proximity of quantum Hall
fillings. In such cases of broken symmetry Wen and Niu have argued 
that the splitting between different states will be exponentially 
small in the area of the system (the number of moves it takes to 
convert one ground state into another)
instead of in the linear dimension, as would be expected from
tunneling processes involving quasiparticles that would move
the system between different quantum Hall ground states. Hence
it would take either a direct examination of the states or
a study of the magnitude of the splitting to decide whether the
ground state cluster is due to topological ordering or merely
a broken translational symmetry (on a manifold of fixed genus,
such as the torus which is what one is likely to study in practice). 
An alternative approach would be to explicitly lift any degeneracies 
due to broken symmetries
by the action of small fields. Any residual degeneracy would
then be topological in origin. For instance the application of
a commensurate periodic potential would lead to the selection
of a unique state in the Wigner crystal phase while it would
reveal the underlying degeneracy of the quantum Hall state
in the case of a quasiparticle Wigner crystal. (We note such
a procedure would seem logically necessary for the topological
degeneracy to track the off-diagonal long range order which
survives in the quasiparticle Wigner crystal if its phonons
are stiff enough.) Similarly in $SU(2)$ quantum Hall ferromagnets
the introduction of a Zeeman term  would be required. Our
discussion here illustrates two more aspects of topological
ordering in quantum Hall systems\\
d) an exponentially small splitting with linear dimension can
be attributed to the presence of fractionalized quasiparticles
that can tunnel across a loop and recombine to move the system
to a new ground state\\
e) it is necessary to break all standard, additional, broken
symmetries explicitly to reveal the underlying topological
degeneracy.

Perhaps the ``cleanest'' as well as the most realistic way to
single out the topological degeneracy is to include the effect
of disorder and thereby examine a quantum Hall phase of finite
extent.
Wen and Niu\cite{wenniu} have offered arguments that the inclusion
of disorder splits the degeneracy by an amount that is $O(e^{-L/\xi})$
in the linear dimension $L$ of the system, $\xi$ being a disorder
correlation length. Their analysis, which holds exactly at $\nu=1/q$
neglects any creation of quasiparticles in the ground state itself,
i.e. the quasiparticle spectrum is assumed to remain gapped in
the presence of disorder. In general this will not be true, and
certainly away from $\nu=1/q$ there will be localized quasiparticles
in the ground state that will give rise to a gapless spectrum.
In a finite volume, this spectrum will acquire a gap that is at
worst polynomially small in $L$ and so if the exponentially
small splitting of the ground states remains, they should not
prevent an identification of the ground state cluster. We do
not know of any detailed examination of whether the ground state
splitting continues to be exponential in this limit -- it would
appear that one cannot merely argue by continuity from the
gapped case due to the singular closing of the gap {\it en route}. 
Neither
is it clear that the operators that move us between states in the 
clean case will continue to work with
randomly localized (or even crystallized) quasiparticles -- here 
again we do not know if a generalization is possible.
Finally, we note that insulating states in the disordered system
are expected to exhibit unique ground states that are separated from 
excited states by, at best, polynomially small gaps coming from
localized electrons.

\subsection{Topological Order in IGTs?} 

The general idea of SF is as
follows (Ref.~\onlinecite{sentfish}b,c). 
The deconfined phase posseses Ising vortex excitations
(visons) that cost a finite amount of energy. As in any
gauge theory where such excitations are possible, in a multiply
connected geometry these can be placed so that their cores 
inhabit the holes and we can expect these configurations to
be long lived, and in an appropriate order of limits they should
be truly meta-stable (that is to say,
infinitely long lived local minima).
For the purposes of the experiment proposed by SF (see below) 
this is sufficient. To make contact with the notion of topological
order, SF wish to relate the presence of visons threading 
holes to an infinite volume limit ground state degeneracy of
$2^h$ on a manifold with $h$ holes, that can
be interpreted as the breaking of a topological symmetry. In the
following we explore this idea in some detail with cylinders
and tori as the manifolds of interest -- going beyond those in genus
while retaining a lattice is tricky, especially when the gauge
theory arises as an effective theory and so we will not venture that
far afield. We begin with pure gauge theories. 

\noindent
{\bf Gauge fields alone:}
As noted by SF, the even IGT on the cylinder at the point $\Gamma=0$ 
exemplifies their ideas. There are two exactly degenerate states, 
which can be written in the $\sigma^z$ representation if one does not 
worry about the constraint. These states have two features of note:
a) that they exhibit a well defined topological 
flux $\hat{F}_z= \prod_\circ
 \hs^z$, where
the product $\prod_\circ$\ 
is taken around the circumference of the cylinder. $\hat{F}_z$\
takes the values $\pm 1$ in the no-vison/vison states;
and b) that there exists an operator $\hat{F}_x = \prod_= \hs^x$ 
where the product $\prod_=$\
is taken along a seam of links, with the seam running along the axis 
of the cylinder
(see SF for details). $\hat{F}_x$
commutes with the Hamiltonian and converts one of the states to 
the other. These two operators capture the two ways of looking 
at the degeneracy, either as a consequence of Ising flux or that
of breaking a ``topological symmetry'' in which a global operator
ceases to annihilate the vacuum. At issue is whether these 
generalize beyond this special point and to Ising gauge fields 
coupled to matter, especially in the QDM limit.

Sticking with the even IGT for the moment, we note that the 
degeneracy is {\it exact} for $\Gamma \ll K$ in perturbation
theory, for a cylinder of finite width. In contrast, it is
clear in the opposite limit $\Gamma \gg K$ that there is 
a unique ground state. This implies that even for a finite width 
cylinder there is a true phase transition {\it en route}.
We note that $\hat{F}_x$ commutes with $H$ at all values of $\Gamma/K$
and that $\hat{F}_z$ is a natural order parameter for this transition, 
being odd under the action of $\hat{F}_x$. Hence the distinction between
the two phases is indeed captured by the action of $\hat{F}_x$ and 
by the development of an Ising flux. We should note though that
$\hat{F}_z$ is measurable only for finite cylinders; being a Wilson
loop, it goes to zero exponentially in (at least) 
the width of the cylinder
at any $\Gamma \ne 0$.

These observations should not really surprise for they involve
a system that is infinite in two space-time directions and finite
in one and hence are equivalent to those concerning the two
dimensional IGT at finite temperature. Such a theory indeed posesses
a phase transition in which the Polyakov loop (a Wilson loop taken
in the time direction) develops an expection value. In the dual
representation this is simply the d=2 Ising phase transition in
a d=3 system that is finite in one 
direction.\cite{svetitskyphysrep}

We return now to the question of working explicitly with gauge
invariant states, \ie\ those that satisfy the local constraint
exactly. Given a state $\left| \Psi \right>$ in the $\sigma^z$
representation, we can construct a state ${\cal P} 
\left|\Psi \right>$
that is gauge invariant by the action of the projector
$$ {\cal P} = \prod_{i} (1/2)[\hat{G}_{IGT}(i) +1]$$
which commutes with the Hamiltonian. Evidently, all gauge invariant
observables have the same value before and after the projection.
While this indicates that our earlier description is correct, it
hides a subtlety of some interest in making contact with earlier
work on the topology of RVB states. To uncover this, note that a
state written explicitly in the $\sigma^x$ basis is automatically
gauge invariant if it involves only even numbers of dimers at each site.
All such configurations can be classified by winding numbers -- one
simply asks how many loops of dimers cross a fixed line bisecting
a set of horizontal bonds. For a finite height cylinder, this number
is either odd and even and the action of the Hamiltonian preserves
this number. Hence the true ground states must be purely even or odd.
Now the vison and no-vison states, when projected, contain both
sectors -- they
correspond to taking the linear combinations 
$\left|{\rm even} \right>\pm \left|{\rm odd} \right>$.
Hence, although they correspond to a different choice of basis in 
the space of the two degenerate states, it is clear that the physical
choice for the standard Hamiltonian is that of purely even and odd 
states which were what were invoked in earlier studies of RVB states.
On the other hand, if we were to allow Wilson loops of
arbitrary length in the Hamiltonian (but with exponentially suppressed 
coefficients to preserve effective locality) we would mix these
states and obtain the vison/no-vison linear combinations split by
an amount exponentially small in the cylinder circumference. In this
case what description one would take to be the correct topological
decomposition in the infinite volume limit would appear to be a matter 
of taste.

On the torus, there is no true phase transition even for the standard
IGT. Instead we find an exponentially small splitting between four
states when the linear dimension $L$ is increased at a fixed coupling
corresponding to the deconfined phase and a splitting of $O(L^0)$
between a unique ground state and the first excited state 
in the confining phase. In terms of the winding number analysis, this
corresponds to the four different combinations of even and odd in
either direction. 

We turn now to the case of the pure odd IGT. Here it is instructive
to work in the $\sigma^x$ representation. By means of the standard
device of using a transition graph between a given state and
a reference state,\cite{Rokhsar88}
 one can again assign a conserved even/odd winding 
number to each configuration. For odd height cylinders, a horizontal
translation by one lattice constant interchanges the two sectors.
Assuming that odd and even height cylinders converge to the same
infinite height limit, it follows then that the ground state must be 
at least twofold degenerate at {\it all} $\Gamma/K$ for infinite height 
cylinders. 
As the $\Gamma \ll K$ analysis in the $\sigma^z$ representation is 
identical to that of the even IGT
except for a different choice of projector, 
\bea
{\cal P} = \prod_{i} (1/2) [\hat{G}_{IGT}(i) - 1] \ ,
\eea 
there is a twofold 
degeneracy in that region. Unlike in the case of the even IGT,
there is a large degeneracy in the extreme opposite limit, $K=0$,
where any dimer covering of the cylinder is a ground state. For
infinite width cylinders, i.e. in the two-dimensional limit, there
is a four-fold crystalline degeneracy as noted earlier. How this
degeneracy is modified by finite cylinder widths is not clear to
us at this point. A preliminary analysis of the QDM on cylinders
indicates that it will exhibit a two-fold degenerate liquid phase 
that does not break any symmetries
as well as a two-fold degenerate 
columnar phase in which the columns run along the cylinder axis. 
Consequently at different cylinder widths the $K/ 
\Gamma\rightarrow 0$
limit may behave differently. We expect that the large circumference
limit will be characterized by symmetry breaking which may either
preserve the two-fold degeneracy of the lowest lying cluster (the
case if the ground state remains liquid for all finite widths)
or increase it by a further factor of two (the case if the ground
state becomes columnar already or finite widths).
In the former case one would have to examine the nature
of the degenerate states to decide what phase they correspond to.

On the torus the deconfined phase has again a four-fold low lying
cluster with a splitting of $O(e^{-L})$ while the confining phase will 
exhibit a cluster of four low lying states with a splitting of
$O(e^{-L^2})$, corresponding to the neccessity of
altering the state over its entire volume instead of just along
a line in the liquid case. (It is worth noting that our previous
argument about translations implementing winding number sector
changes implies that there is an exact two fold degeneracy due
to translational symmetry breaking on odd by even tori.)
So on the torus one would need to
examine the size dependence of the splitting or the correlations
in the ground states to distinguish the two four-fold degeneracies
from each other. Alternately, as in the quantum Hall case one could
turn on symmetry breaking fields that would lift the degeneracy
in the crystalline phase but not in the liquid, deconfined, phase.
We note that in the context of the cuprates, this is the
case of maximum interest.

To summarize: The behavior of Ising gauge fields alone does display
a ``family resemblance'' to the quantum Hall case with regards
to points a,b,d and e made earlier. With respect to c the fundamentally
discrete character of this problem makes it unlikely that there
is an analog. That being said we should note that in the QDM limit
it does not really go beyond the previous analysis of RVB wavefunctions
in terms of winding number sectors -- the latter is an analysis in
terms of electric fluxes (the momenta conjugate to the gauge fields).

In this regard the really interesting claim of SF is that the
phase obtained at finite doping is {\it also} characterized by
topological order. As the even/odd classification breaks down upon
doping, this would be a feature not obtained by the previous
analysis. In the language of the IGT we must ask what
happens when we add matter to the problem.

\noindent
{\bf Gauge fields with matter:} 
We note at the outset that this might be expected to differ from
the quantum Hall case. In the latter the states differ, in a sense,
by the insertion of integer numbers of flux quanta through the
holes. By contrast in the IGT problem, the vison will be seen by
matter fields as {\it half} a flux quantum.

Nevertheless, the effect of the additional flux can be 
exponentially attenuated if the matter fields are gapped on
their own. The simplest such case is that of the even IGT with
Ising matter. While $\hat{F}_x$ no longer commutes with $H$,
perturbative considerations indicate that in the deconfined phase
there are two low lying states with a splitting that is $O(e^{-L})$\
at large $L$, which goes away on leaving this phase. So in this case
it is indeed possible to relate the deconfined phase to a two-fold
degeneracy. Having identified the two ``ground'' states, one can test
them for the presence of flux. With matter present, the even and odd
sectors are now connected and the states will exhibit (small)
expectation values of the Wilson loop consistent with the presence and
absence of a vison. 

One might wonder if it is possible to relate the two low lying states 
by the action of $\hat{F}_x$.  It turns out that the attempt to create 
one from the other by its action will yield a vanishing overlap in the 
limit of infinitely long cylinders. 
This result can be obtained perturbatively near the trivial 
point $\Gamma/K=0$, $u/\lambda=0$. At this point, the ground state
with $F_z=0$, $\left|\Phi_0\right>$\ has $\sigma^z\equiv1$\ and
$\tau^x\equiv-1$, whereas the state
$\left|\Phi_1\right>\equiv\hat{F}_x\left|\Phi_0\right>$\ differs in
that the horizontal $\sigma^z$\ are flipped along one seam along the
axis of the cylinder of height $H$\ that the lattice resides
on. Carrying out perturbation theory to second order in $u/\lambda$\
yields the perturbed wavefunctions
$\left|\Phi_0^{2}\right>$\ and $\left|\Phi_1^{2}\right>$,
respectively.  One then finds 
\bea
\left<\Phi_1^{2}\right|\hat{F}_x\left|\Phi_0^{2}\right>=
\exp(-2H(u/\lambda)^2).
\eea
where we have exponentiated the linear answer that perturbation
theory actually produces. General random walk arguments indicate
that the exponential dependence on the height is exact though
the coefficient will be modified at higher orders in perturbation
theory.
In sum, the degeneracy is recovered in the infinite system size limit 
but the topological symmetry operation no longer takes us between
ground states. 

The case of greatest interest is that of charged matter coupled to 
Ising gauge fields. SF have suggested that spinon and holon
fields coupled to an Ising gauge field are the correct low energy theory
of a variety of strongly correlated systems and have argued that
anomalous non-superconducting phases would be charaterized by
topological degeneracies that could, in principle, be used to
search for such phases in numerical studies or variational studies.

In the QDM framework, we are concerned with adding holons to a dimer 
liquid. If the dimers remain liquid, then we have a doped phase that
might be expected to inherit topological degeneracies from the parent
insulating state. It would appear that there are three possibilities: \\
a) the holons localize \\ 
b) the holons are bosonic and condense thereby giving rise to 
a superconductor\\
c) the holons are fermionic and produce a gapless 
spectrum.\cite{readcha}

In case a) one has perhaps the strongest argument for a surviving
topological classification and associated degeneracy. Certainly if 
the holons are truly immobile, one can define even and odd sectors 
for that given configuration. If they are localize on some length 
scale, the classification is no longer strict but it seems plausible
that for system sizes much bigger than their localization length,
the degeneracy is recovered.

In case b) the system ends up with a superconducting vortex threading it 
and so the question is moot. 

In case c) we would truly have a non-Fermi liquid but metallic phase.
Unfortunately in such a system it would appear that all gaps are
polynomially small and so it will not be possible to select a
ground state multiplet in an operational sense. From the point of 
view of the QDM, all states involve holons and dimers in correlated
motion around the torus and no topological character is evident.
As we were unable to construct use the topological symmetry operator
of the pure gauge theory in the case of Ising matter, we will not 
succeed here either.\cite{fn-singlehole}
It would appear then that in this case the
non-Fermi liquid character will not give rise to a meaningful
topological degeneracy.

\subsection{Flux trapping experiments}

We are however, still left with the possibility that the states
of the doped QDM are characterized by finite (if exponentially
small) Ising flux measured by the Wilson loop. If such a state
has a net vison content in a non-superconducting phase, it
would seem likely that it would nucleate a vortex if the
parameters are changed to condense the holons. This would then
realize the SF scenario.\cite{fn-meta}

From our considerations in the last section, we conclude that
a flux trapping experiment that cycles between phases with
the holons localized and then superconducting would be the
most robust while that between the latter and the strange
metal is hard to predict without a more detailed theory of
the metal. In either case, the issue appears quite delicate
from the QDM viewpoint, in which the system is required to
remember rather delicate phase relationships between different
components as the parameters change. Of course one of the
strengths of the vison viewpoint is, that by focussing attention
on the relevant collective co-ordinate, it suggests that this
is an artefact of looking too microsocpically. Further studies
of the doped dimer model could be very instructive in this
regard.

\section{Discussion}

In this paper we have established and discussed several important 
connections existing between short-range RVB phases, quantum dimer
models, and Ising gauge theories, which have significant implications for the 
problem of spin-charge separation in strongly correlated systems.

To begin with, we showed that there exists a natural
physical interpretation of the Hilbert space of RVB phases, and that its 
Ising character follows  directly from the nature of the states 
themselves: short-ranged RVB states are
naturally described in terms of short range spin singlets which are
either present or absent. Thus, from the point of view of the space of
states, a description of the dimer Hilbert space should have a natural description in
terms of Ising variables living on the links of the lattice. As a naive
description of this form is seriously overcomplete, it is clearly necessary
to impose constraints at each site which then generate a family
of local gauge transformations that leave the Hamiltonian invariant. An Ising
constraint would be sensitive only to the number of valence bonds modulo two.
However, since the number of valence bonds  (dimers) is
conserved, the effective Hamiltonians associated with these states must have a 
natural
local conservation law and consequently  a local $U(1)$ symmetry, instead of the
$Z_2$ ``natural" symmetry  of an Ising Hilbert space. 
We further showed that quantum dimer models can indeed  be
realized as (odd) Ising gauge theories with additional couplings which project out
forbidden configurations of dimers (valence bonds). Thus, while the
Ising {\sl variables} provide a natural and economical description of
the Hilbert space, the native symmetry to the physics of short-ranged
RVB states is
actually $U(1)$ and not $Z_2$.  

However, we also found that the phase
structure of generalized quantum dimer models depends on how the local $U(1)$ symmetry
is realized. Superficially, a $U(1)$ gauge symmetry may seem to rule out deconfined
phases since it is quite well known that the vacuum sector of $U(1)$
gauge theories are confining in $2+1$ dimensions. It turns out
that for the case of the gauge theoretic description of quantum dimer
models the situation is more subtle. For instance, on the square lattice
the ground state turns out to be confining, and thus it is not a spin
liquid. In contrast, on non-bipartite lattices the situation is quite 
different. Indeed in such cases dimers connecting sites on the same sublattice 
give rise to  matter fields that carry two units of the
$U(1)$ gauge charge.  In this case the deconfinement mechanism of 
Ref.\ \onlinecite{fradshen} (derived for the even IGT) can be expected to apply 
and  both a confining and a deconfined phase may exist. In the 
deconfined phase the effective remaining ``unbroken" local symmetry is reduced 
to $Z_2$.
Thus, this mechanism of spin-charge separation relies entirely on the existence 
of a deconfined phase in the Ising gauge theory.
A local $Z_2$ symmetry is also central to the work of SF \cite{sentfish} although 
their point of departure is a superconducting state with Cooper pairs. We
have noted that their starting Hamiltonian has more degrees of freedom than
the single band $t-J$ type models that we have in mind so their identification
of the Ising variable is not as microscopic. However valence bonds are
sufficiently akin to Cooper pairs \cite{pwa87} that one is tempted to guess
that both approaches describe the same physics.

The considerations presented above assume that the confinement-deconfinement
structure of the phase diagram  of {\sl even} Ising gauge theories holds also 
for the {\sl odd} Ising gauge theories. Although this is not rigorously 
established, there is substantial evidence, including the results reported in
this paper, that the main difference bewteen even and odd theories is to 
associate confinement with phases in which translation and/or rotational 
invariance are spontaneously broken, such as valence bond crystals and stripe 
states. In contrast, deconfined phases are always liquids. The exception
to this is the case of $d=1$. Here the even IGT, whose ground state is
translationally invariant, confines at all couplings while the odd IGT whose
ground state breaks translational symmetry, and hence would be expected to
be confining by our previous remarks, allows test charges to be
separated at a finite cost in energy. This peculiar feature is, of course,
the topological mechanism of spin-charge separation in $d=1$ wherein the
charges are accomodated on a pair of solitons interpolating between
the two ground states.

A conclusion that emerges from this line of argument, is that there
is a fundamental difference behind the mechanism of spin-charge
separation in one-dimensional and two-dimensional systems.
Indeed, in one dimension holons and spinons are actually {\sl topological
solitons}, and spin-charge separation is a topological phenomenon, peculiar to 
the kinematics of one-dimensional systems. In contrast, in two dimensions 
(and higher) spin-charge separation
relies on the existence of {\sl deconfinement} in the sense of liquidity, which 
is a property of the spectrum of states in a particular {\sl phase} of matter, 
and as such it does not hold in general; deconfinement takes place in some 
cases, such as the triangular lattice which can have a spin liquid ground 
state,\cite{MStrirvb} whereas confinement is naturally realized on
the square lattice.\cite{fradkiv,sachvojt} 

The question of the existence of a deconfinement mechanism of gauge theories with 
dynamical
matter at finite density has a long history in high energy physics which which is 
rather similar
to the quest for a spin-charge separated state in condensed matter physics. The
difficulties of defining order parameters and other tests of confinement has been 
a central theme
in that field since the late seventies. In fact it has long been recognized in that 
field that no
such tests can exist in terms of gauge invariant local operators (such as order 
parameters) or
Wilson loops, if the dynamical matter fields carry the fundamental gauge charge. 
A related and
important current question is if hadronic matter at finite density is generally and 
smoothly
connected to conventional nuclear matter, or if a genuine quark-gluon plasma exists 
as a state of
matter with unique measurable signatures. This latter phase is indeed precisely 
the equivalent of
the spin-charge separated phase discussed here.

Finally, we have also discussed the question of topological degeneracy of the 
deconfined spin liquid states, and their possible detection which we argue
is not contingent upon the former in any precise sense. 
We have discussed in some detail the set of {\it desiderata} associated
with the notion of a topological degeneracy by
reviewing the case of clean quantum Hall systems at the ideal filling
fractions. We have discussed the applicability of these to disordered
quantum Hall systems and then to the case of Ising gauge theories.
We find that while there is certainly a sense in which IGTs in their
deconfined phases exhibit a finite ground state degeneracy in the 
thermodynamic limit, in general there is no accessible operational test for 
this degeneracy short of a full solution of
the spectrum of states. In particular we find that the overlap of a test state 
with one vison
wrapped around a non-contractible loop is orthogonal to any ground state in the 
thermodynamic
limit, and therefore it does not connect distinct degenerate states. This behavior 
stands in contrast with what happens in ideal quantum Hall fluids and chiral spin 
states, although it may be generic in more realistic cases.

\section*{Acknowledgements} We are grateful to 
T.\ Senthil and M.\ P.\ A.\ Fisher
for several discussions on the the connections between QDMs and IGTs
and especially on their work on topological order, as well as
for comments on the manuscript. Several of the
results on the connections between GDMs and IGTs have also been
derived independently by them. We are also grateful to D. Huse,
S.\ Kivelson, S.\ Sachdev and N.\ Read for valuable discussions. EF thanks
S.\ Kivelson for a prior collaboration on
closely related questions, and RM and SLS would similarly like to
thank P.\ Chandra for collaboration on related work.  This work was supported
in part by grants from the Deutsche Forschungsgemeinschaft, the NSF
grant Nos. DMR-9978074 (Princeton) and DMR-9817941 (UIUC), the A.\ P.\ Sloan 
Foundation and the David and Lucille Packard Foundation.

\appendix
\section{Duality of IGT\lowercase{s} 
with Ising models in $\lowercase{d}=2+1$}
\label{app:dual}
We show that the GDM with the Hamiltonian given by $ \hat{H}_{GDM}$\
(Eq.~\ref{eq:HIGT})\ in $d=2+1$\ is dual to an Ising model with the
Hamiltonian:
\bea
G=+1:\ && \ H_+=-k\sum_-\hS^z_i\hS^z_j+\gamma\sum_.\hS^x_i\ ,\\
G=-1:\ && \ H_-=-\sum_-k_{ij}\hS^z_i\hS^z_j+\gamma\sum_.\hS^x_i\ ,
\eea
where the sums $\sum_-$\ run over the links of the dual 
lattice and the $\sum_.$\ over the sites. The $\hS$\ are Pauli
spin operators defined on sites of the dual lattice, $k>0$\ 
and $|k_{ij}|=k$. The case of $G=-1$\ is known as a fully frustrated 
Ising
model (FFIM) since each plaquette $\Box$\ has to have at least one 
frustrated bond: $\prod_\Box(k_{ij}/k)=-1$, whereas the case $G=+1$\ 
is a ferromagnetic Ising model (FIM).

The starting point of the duality is the identification of a
frustrated bond in the Ising model
with a dimer in the GDM. One can easily
convince oneself that each plaquette in the FIM (FFIM) has to have an
even (odd) number of frustrated bonds, which takes care of the
constraint $G=+1\ (-1)$. 

Conversely, each dimer
state corresponds to a unique spin state (up to a global Ising
reversal). This can be seen by taking a reference spin configuration,
for example $\hS^z\equiv1$, which corresponds to a reference
configurations of dimers, namely one without dimers ($G=+1$), or to a
columnar dimer state ($G=-1$). Any other dimer configuration can then
be used to generate a `transition graph' (see
Ref.~\onlinecite{Rokhsar88}), obtained by superimposing that dimer
configuration with the reference dimer configuration. The resulting
transition graph contains only closed loops. To fix the overall Ising
redundancy, an arbitrary reference spin is chosen to point 
up.\cite{fn-block}
The
orientation of any other spin is then obtained by counting the number
of dimers in the transition graph any line connecting that spin to the
reference spin crosses. If this number is even, the spins are aligned,
otherwise they are antialigned.

To construct the equivalence between the Hamiltonians, we only need to
note two facts. Firstly, the presence of a satisfied bond gains an
energy $k$, whereas a frustrated bond costs the same amount of
energy. Translating this into a statement about the absence, presence
of a dimer, we obtain the identification $\Gamma=k$\ for the first
pair of coupling constants. Secondly, note that flipping a spin $S_i$\
implies exchanging all its satisfied bonds for frustrated ones and
vice versa. This is equivalent to exchanging occupied and empty dimer
links of the plaquette $i$, at the center of which $S_i$\ is located.
This immediately yields the identification of the spin flip effected
by the $\hS^x$\ operator with the plaquette term in $\hat{H}_{GDM}$,
together with $\gamma=\kappa$. This completes the demonstration of
duality.

\section{The Polyakov loop term in the action}
\label{app:berry}

In order to see the connection between the constraint in the
Hamiltonian formalism and the role of the Polyakov loop
in the path integral,\cite{polloop,sussconst} it is useful
first to recollect the appropriate construction for the electromagnetic
gauge field. The
Lagrangian density for the free electromagnetic field is
\bea
{\cal L}[A,j]=\frac{1}{2} \left({\vec E}^2-{\vec B}^2\right)-A_0 j_0
\nonumber
\eea
where
\bea
E_i&=&\partial_0 A_i+\partial_i A_0
\nonumber\\
B_i&=&\epsilon_{ijk}\partial_j A_k
\nonumber\eea
and $j_0$ is a static charge distribution, say
\bea
j_0(z)=\delta(z-x)-\delta(z-y)
\nonumber\eea
for two static charges at $z=x,y$ (with charge $\pm 1$ respectively).
The path integral in $D$ space-time dimensions is
\bea
Z[j]=&&\int {\cal D} A_\mu \;\; e^{\displaystyle{i\int d^Dx \; {\cal L}[A,j]}}
\nonumber\\
=&& \int {\cal D} A_\mu\;\; e^{\displaystyle{i\int d^Dx \left[ {\cal L}[A,0]-A_0 
j_0\right]}}
\nonumber\eea
Thus, 
\bea
\frac{Z[j]}{Z[0]}=
\langle 
e^{\displaystyle{-i\int dx_0 A_0({\vec x},x_0)}}
e^{\displaystyle{+i\int dx_0 A_0({\vec y},x_0)}}
\rangle
\nonumber\eea
namely, the expectation value of the product of two Polyakov loops.

It is easy to show that in the Hamiltonian picture the Polyakov loops
become static sources in the Gauss' Law constraint.\cite{polloop} Let
us rewrite the path integral by using the coherent state
representation, which is a integral over both the vector potential
$A_i$, the conjugate momenta, the electric field $E_i$, and the
Lagrange multiplier field $A_0$. Glossing over issues
related to gauge fixing, gauge copies, and Faddeev-Popov determinants,
one writes, 
\bea Z[j]=\int {\cal D}E_i {\cal D}A_i {\cal D}A_0 \;\; 
e^{\displaystyle{i \int d^Dx \; {\cal L}[A_i,E_i,A_0,j]}} \nonumber
\eea 
where 
\bea 
{\cal L}[A_i,E_i,A_0,j]=-E_i &&
\partial_0 A_i -\frac{1}{2} \left({\vec E}^2+{\vec B }^2\right)\nonumber \\
&& + A_0
\left( \partial_i E_i -j_0\right) \nonumber\eea
Thus, we see that the role of
$A_0$ is of a Lagrange multiplier that forces Gauss' Law 
\bea
\left[{\vec \nabla} \cdot {\vec E}-j_0\right]|{\rm Phys}\rangle=0 
\nonumber
\eea
as a constraint on the physical Hilbert space. Thus, the Polyakov
loops are equiv alent to static sources. Notice that this is really
the Hamiltonian picture since we get that the momentum canonically
conjugate to $\vec A$ is $-{\vec E}$, as we should. 

We now turn to the case of the Ising gauge theory. Consider the
Hamiltonian of Eq.~\ref{eq:HIGT}. For convenience, we define
$\hh{\mu}^x=(1-\hs^x)/2$, so that the Hamiltonian is written, up to a
constant,
 \beq \hat{H}_{GDM}=
2\Gamma\sum_{-}\hh{\mu}^x-\kappa\sum_\Box\hs^z\hs^z\hs^z\hs^z.
\label{eq:HMGT}
\eeq 

It turns out to be convenient to rewrite the constraint operator
$\hat{G}=\hat{\tau}^x\prod_+\hsi^x$\ as follows. 
Let $\hat{L}(i)=\hh{\nu}(i)^x+\sum_+\hh{\mu}^x$, where 
$\hh{\nu}^x(i)=(1-\hh{\tau}^x(i))/2$.
Then the projector enforcing
$\hat{G}(i)\phy=\Upsilon(i)\phy$\ at site $i$\ is 
given by
\bea
\hP(i)=(1/2)\left[ 1+(-1)^{(\hh{L}(i)+\xi(i))}\right]\ .
\eea
Here, $\xi(i)=(1+\Upsilon(i))/2$\ is $0$ for an even site and $-1$
for an 
odd site. In the absence of matter, all sites of the even (odd) 
theory are even (odd), but the addition of a hole at a site changes
it from even to odd and vice versa, so that the following treatment 
is appropriate for static matter. 

We now Trotterise the partition function $Z(\beta)$\ at temperature
$1/\beta$ and obtain
\bea
Z(\beta)&=&\Tr\left( \exp(-\beta \hat{H})\hP\right)\nonumber\\
&=&\lim_{\epsilon\rightarrow0}\prod_{\zeta=0}^{N-1} 
\bra{\left\{\hsi_{\zeta+1}^z \right\}} 
\exp(-\epsilon \hat{H})\hP
\ket{\left\{\hsi_{\zeta}^z \right\}}
\, ,
\label{eq:zbeta}
\eea
where the Greek letter $\zeta$\ labels the (imaginary) time slices, 
and $\epsilon=\beta/N$, and the $\hsi^z$\ are eigenstates of $\hs^z$.

Consider a single term in the product, which we evaluate by inserting
a complete set of eigenstates of $\hs^x$:
\bea
\bra{\left\{\hsi_{\zeta+1}^z \right\}} 
&&\exp(-\epsilon \hat{H})\hP
\ket{\left\{\hsi_{\zeta}^z \right\}}
=
\Tr_{\left\{\hsi_{\zeta}^x \right\}}
\sum_{\{\lambda_\zeta(i)=0,1\}}(1/2)^{N_s}\nonumber\\
&&\exp\left(-\epsilon \kappa \sum_\Box\hsi^z\hsi^z\hsi^z\hsi^z
\right)
\exp\left(-2\epsilon \Gamma\sum_-\mu^x
\right)
\times
\nonumber\\
&&\prod_i\exp\left[i\pi\lambda_\zeta(i)
\left(\sum_+\mu^x+\xi(i)
\right)
\right]
\times
\nonumber\\
&&\braa{\left\{\hsi_{\zeta+1}^z \right\}} 
\ket{\left\{\hsi_{\zeta}^x \right\}} 
\braa{\left\{\hsi_{\zeta}^x \right\}} 
\ket{\left\{\hsi_{\zeta}^z \right\}} 
\nonumber
\eea
Here, $N_s$\ is the number of sites, and $N_b$\ the number of bonds. 
We have rewritten the projector
as an exponential and turned the operators into numbers by letting
them act on their appropriate eigenstates. Note that 
$\braa{\{\hsi_{\zeta}^x \}} 
\ket{\{\hsi_{\zeta}^z \}} =
2^{-N_b/2}\exp\left(i\pi\sum_-\mu^x_\zeta\nu^z_\zeta\right)$, 
where the sum runs over all links in timeslice $\zeta$.

Collecting together the terms involving the $\hsi^x$, we obtain
\bea
\sum_{\left\{\hsi_{\zeta}^x \right\}}\prod_{\iD}\frac{1}{2}&&
\exp\left\{
\mu^x_\zeta(\iD)
\left[
-2\epsilon\Gamma+
i\pi
\left(\lambda_\zeta(i)+\lambda_\zeta(i+\cD)+
\right.
\right.
\right.
\nonumber\\
&&\left.\left.\left.
\mu^z_\zeta(\iD)+\mu^z_{\zeta+1}(\iD)
\right)
\right]
\right\}\, .
\eea
Here, $\iD$\ labels the bond connecting site $i$\ with its neighbour
in a spatial direction labelled by $\cD$. 

The term in parentheses in the 
foregoing equation can be turned into a plaquette product by defining
$\hsi^z=(1+\lambda)/2$\ on the temporal links, so that this expression
becomes
\bea
\prod_{\iD}\frac{1}{2}\left\{
1+\exp(2\epsilon\Gamma)\prod_\Box\hsi^z
\right\}= \prod_{\iD} \frac{\exp(K^\tau\prod_\Box\hsi^z)}{2\cosh K^\tau}\, .
\eea
In the last step, we have used the fact the product over plaquettes 
containing temporal bonds occurring in this expression, 
$\prod_\Box\hsi^z$, can only take on values $\pm1$. The equality
holds for a temporal coupling $K^\tau=\tanh(-2\epsilon\Gamma)$.

Putting this result back into Eq.~\ref{eq:zbeta}, using $2\cosh
K^\tau\rightarrow \exp(K^\tau)$\ for $\epsilon\rightarrow0$, and 
substituting for $\lambda$\ in terms of temporal $\hsi^z$,
we obtain
\bea
Z(\beta)&=&(1/2)^{N_s}\Tr_{\left\{\hsi\right\}} 
\left[\prod_|\hsi(i)^{\xi(i)}\right]
\times\nonumber\\
&&\exp\left[-K^s\sum_\Box\hsi\hsi\hsi\hsi
-K^\tau\sum_\Box\hsi\hsi\hsi\hsi\right]
\, ,
\eea
where $K^s=\epsilon\kappa$, the first sum in the second line 
runs over
spatial plaquettes, the second over temporal plaquettes.
The trace now runs over all the $\hsi$, 
both in spatial and temporal directions.

Crucially the product $\prod_|$\ runs over the temporal links -- 
this
is the Polyakov loop term. It contributes a nontrivial phase for all
the odd sites. This is what we set out to show.


\begin{references}
\bibitem{fn-subir}
There does, however, exist an approach\cite{sachvojt} 
starting not from a spin liquid
but rather from a columnar valence bond solid state, which upon
doping reaches
a d-wave superconductor via an intermediate regime of stripes.
\bibitem{fn-cd}
SF study an effective theory in the sense that their starting
action contains separate fermionic and superconducting degrees of
freedom.
\bibitem{fn-tetra} The reader will have noticed that, in the
absence of a hardcore constraint, valence bonds on different sites can
join to form not just dimers and tetramers but much more complicated
lattice animals.
\bibitem{fn-singlehole} 
We have examined this possibility explicitly
for the case of the QDM doped with a single hole.
Most simply, consider a $d=1$\ QDM
with an odd number of sites, $L$, and one boson hopping with 
amplitude
$-t$.  The ground state is then the state
$\left|\psi\right>=1/\sqrt{L}\sum_{i=1}^L\left|i\right>$, where
$\left|i\right>$\ denotes the (unique) state with the boson at site
$i$. Its energy is thus $E_\psi/t=-2$, and the energy of the first
excited state, $\left|\chi\right>$, is separated by a gap of
$(E_\chi-E_\psi)/t\propto L^{-2}$.  By contrast, consider the state
$\left|\phi\right>\equiv \hat{F}_x\left|\psi\right>$: 
the expectation value of
its energy is higher than the ground state by
$(\left<\phi\left|H\right|\phi\right> -E_\psi)/t\propto L^{-1}$. 
This argument can be extended to higher $d$, where the exact
single-hole ground state $\left|\Psi\right>$\ and first excited states
also differ in energy by an amount $\propto L^{-2}$. A hole hopping in
the state $\hat{F}_x\left|\Psi\right>$\ will encounter a flipped bond
variable $\sigma^x=-1$\ in $L$\ out of $L^2$\ of its hops, which gives
rise to an energy difference of $L^{-1}$\ compared to the ground
state, as in the case in one dimension.
\end{references}
\end{document}